\documentclass[twocolumn,amsmath,amssymb,prl]{revtex4}
\usepackage{latexsym}
\usepackage{graphicx}
\usepackage{psfig}
\usepackage{color}
\usepackage{verbatim}
\usepackage{multirow}
\usepackage[letterpaper, margin=1in]{geometry}
\usepackage{siunitx}

\begin{document}

\preprint{}
\title{Controlling Fermi level pinning in near-surface InAs quantum wells}
\author{William~M.~Strickland}
\author{Mehdi~Hatefipour}
\author{Dylan~Langone}
\author{S. M. Farzaneh}
\author{Javad~Shabani}

\affiliation{Center for Quantum Phenomena, Department of Physics, New York University, NY 10003, USA}

\date{\today}

\begin{abstract}

Hybrid superconductor-semiconductor heterostructures are a promising platform for quantum devices based on mesoscopic and topological superconductivity. In these structures, a semiconductor must be in close proximity to a superconductor and form an ohmic contact. This can be accommodated in narrow band gap semiconductors such as InAs, where the surface Fermi level is positioned close to the conduction band. In this work, we study the structural properties of near-surface InAs quantum wells and find that surface morphology is closely connected to low-temperature transport, where electron mobility is highly sensitive to the growth temperature of the underlying graded buffer layer. By introducing an In$_{0.81}$Al$_{0.19}$As capping layer, we show that we can modify the surface Fermi level pinning within the first nanometer of the In$_{0.81}$Al$_{0.19}$As thin film. Experimental measurements show a strong agreement with Schrödinger-Poisson calculations of the electron density, suggesting the conduction band energy of the In$_{0.81}$Ga$_{0.19}$As and In$_{0.81}$Al$_{0.19}$As surface is pinned to \SI{40}{\milli eV} and \SI{309}{\milli eV} above the Fermi level respectively.

\end{abstract}

\maketitle

Narrow band gap semiconductors possess a large $g$ factor, small electron effective mass and strong spin-orbit coupling, making them excellent candidate materials for next-generation high-speed electronics \cite{kroemer2004, del_alamo_nanometre-scale_2011} and mesoscopic and topological superconducting circuits \cite{LutchynReview}. In particular, InAs has the unique characteristic of having its surface Fermi level pinned in the conduction band \cite{tsui1970, MeadPR64, bhargava}, forming an ohmic contact with metals such as aluminum. This is particularly relevant in the context of the superconducting proximity effect, where the semiconductor inherits pairing properties of the superconductor \cite{mayer2019, Henri17,  MatthieuPhase2021, Fabrizio17, FornieriNature2019, Ren2019, mayer2019anom}. Advancements in the growth of an epitaxial, superconducting aluminum thin film on an InAs quantum well have led to reports of an abrupt, smooth and uniform interface between the two materials~\cite{Shabani2016, Kaushini2018, sarney2018, sarney_2020_metallization}, enabling ballistic transport between two superconducting leads through the semiconductor weak-link \cite{mayer2019, MortenPRA2017, dartiailh_missing_2021}.

For many device applications, high electron mobility is a stringent requirement. Continual optimization of the growth of InAs quantum well structures has resulted in a steady increase of electron mobility over the years. There have been reports on high mobility InAs two-dimensional electron gases (2DEG) grown on GaAs \cite{bolognesi_interface_1992, nguyen_growth_1993, thomas_buffer-dependent_1997, zverev_magnetotransport_2004, gozu_low_1998, cai_band_2002, Richter1999, mendach_strain_2002}, InP \cite{Hatke, ShabaniAPL2014, Wallart05,nakayama_modulation_1999} and GaSb \cite{shojaei_studies_2015, shojaei2016, ma2017, Tschirky17, thomas2018}. While InAs and GaSb are nearly lattice matched, the commercial availability and ease of nanofabrication of arsenide-based heterostructures have attracted attention to InP or GaAs substrates. The compressive lattice mismatch of 3.3\% between InAs and InP however necessitates the use of a metamorphic, step-graded buffer layer, which relies on the formation of misfit dislocations at layer steps to relax strain on the overlayer. 

While InAs two-dimensional electron gas (2DEG) mobilities in excess of $\SI{1e6}{cm^2/Vs}$ have been reported \cite{Hatke} in deep quantum well structures where the active region is hundreds of nanometers below the surface, transport in near-surface quantum wells suffer gravely from surface scattering \cite{Kaushini2018}. It was found that a thin In$_x$Ga$_{1-x}$As top barrier can separate the electron wave function from the surface, significantly enhancing the mobility \cite{Shabani2016, Kaushini2018}. While the thickness of this layer should be kept relatively thin ($<$10 nm), the composition of indium could tune the surface Fermi level pinning position. There have been few studies however detailing the use of different barrier materials and it's effect on electron transport in near-surface InAs quantum wells~\cite{JoonSue, Kaushini2018, Shabani2016, Pauka2019, Yuan2020, cimpoiasu_charge_2020}.

\begin{figure*}[ht]
    \centering
    \includegraphics[width=\textwidth]{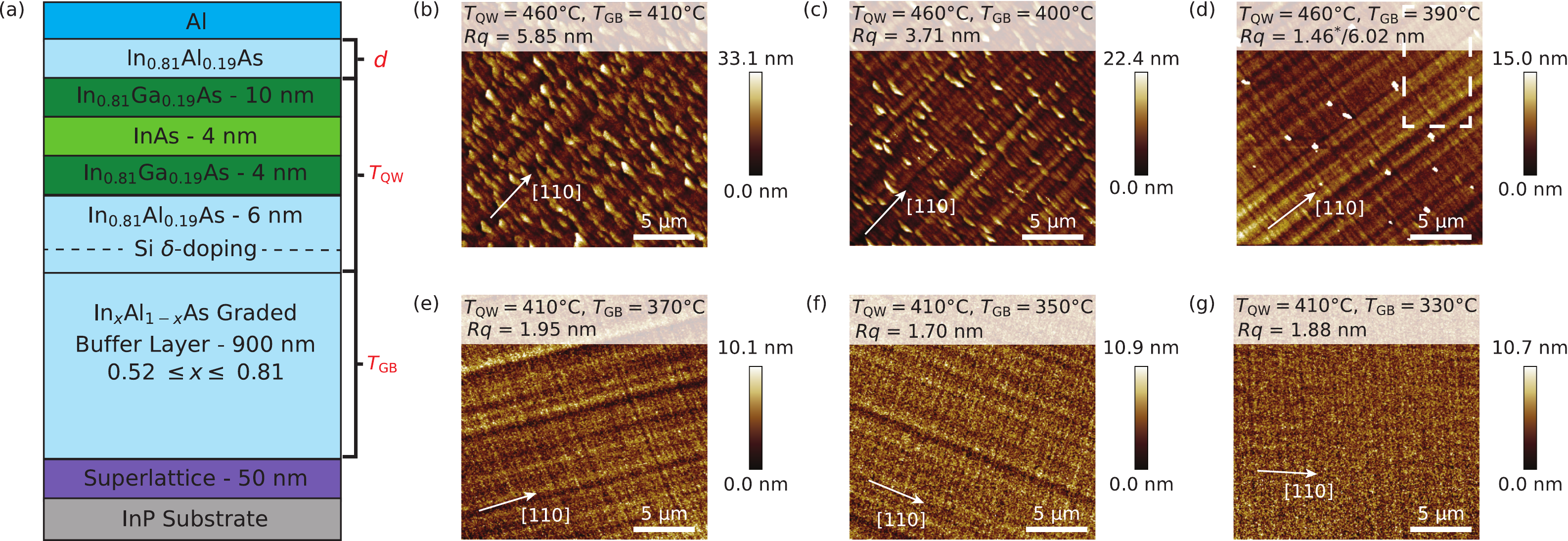}
    \caption{(a) Schematic of the layer structure. $T_\text{GB}$ and $T_\text{QW}$ are the growth temperatures of the graded buffer layer and quantum well, and $d$ is the thickness of the In$_{0.81}$Al$_{0.19}$As barrier layer. (b-g) AFM images of the aluminum surface of samples with varying $T_\text{GB}$ and $T_\text{QW}$. Root-mean-squared surface roughness, $Rq$ and the $[110]$ direction for each are annotated. The growth temperatures are (b) $T_\text{QW}=460\SI{}{\celsius}$ and $T_\text{GB}=410\SI{}{\celsius}$, (c) $T_\text{QW}=460\SI{}{\celsius}$ and $T_\text{GB}=400\SI{}{\celsius}$, (d) $T_\text{QW}=460\SI{}{\celsius}$ and $T_\text{GB}=390\SI{}{\celsius}$, (e) $T_\text{QW}=410\SI{}{\celsius}$ and $T_\text{GB}=370\SI{}{\celsius}$, (f) $T_\text{QW}=410\SI{}{\celsius}$ and $T_\text{GB}=350\SI{}{\celsius}$, and (g) $T_\text{QW}=410\SI{}{\celsius}$ and $T_\text{GB}=330\SI{}{\celsius}$. The asterisk next to the $Rq$ value for Sample C is calculated over the area denoted by the white dashed box, while $Rq$ over the whole area is shown without an asterisk.}
    \label{fig:fig1}
\end{figure*}

\begin{table*}
\centering
\begin{tabular}{|| c | c | c | c | c | c ||} 
 \hline
 Sample & $T_\text{QW}$ (\SI{}{\celsius}) & $T_\text{GB}$ (\SI{}{\celsius}) & $Rq$ (nm) & $n_\text{meas}$ ($\times 10^{11} \SI{}{\centi m}^{-2}$) & $\mu$ (\SI{}{\centi m}$^2$/Vs)\\ [0.5ex] 
 \hline\hline
  A & 460 & 410 & 5.85 & 6.88 & 14900 \\ 
  B & 460 & 400 & 3.71 & 7.46 & 17200 \\ 
  C & 460 & 390 & 1.46*/6.02 & 10.60 & 9410 \\ 
  D & 410 & 370 & 1.95 & 8.91 & 27000 \\ 
  E & 410 & 350 & 1.70 & 6.01 & 19500 \\ 
  F & 410 & 330 & 1.88 & 7.81 & 25700 \\ 
  \hline
\end{tabular}
\caption{Table with growth temperatures $T_\text{GB}$ and $T_\text{QW}$ mobilities $\mu$ and 2D electron densities $n$. Two $Rq$ values for Sample C correspond to $Rq$ calculated in a selected area shown in Fig.~\ref{fig:fig1}(d) denoted by the white dashed box (with an asterisk next to the value), and over the whole area.} 
\label{table:table}
\end{table*}

\begin{figure}
    \centering
    \includegraphics[width=0.45\textwidth]{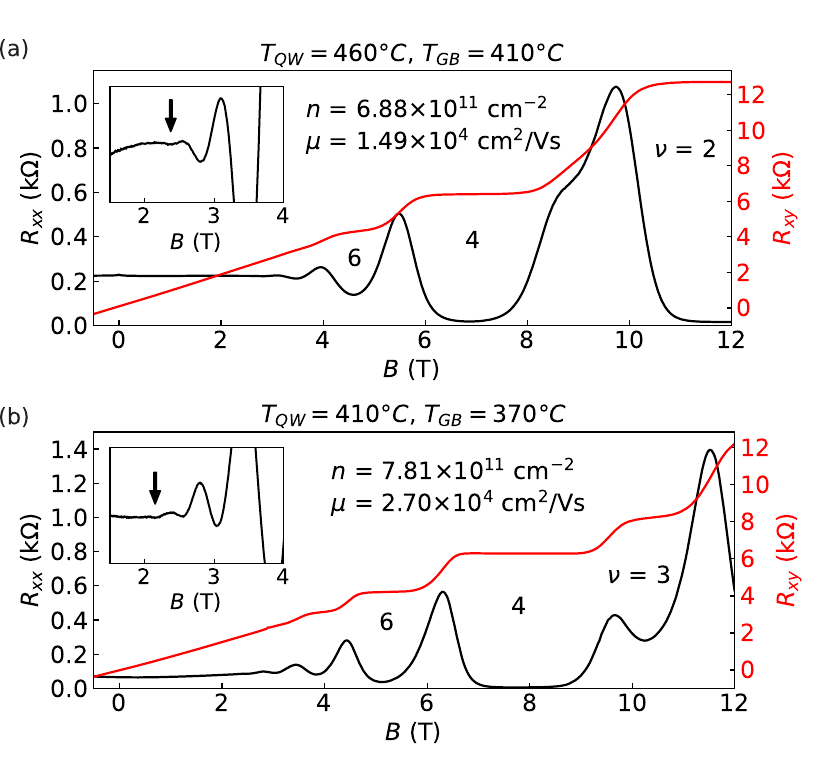}
    \caption{Magnetotransport measurements of the longitudinal resistance $R_{xx}$ (black, left) and Hall resistance $R_{xy}$ (red, right) as a function of magnetic field $B$ for (a) samples A, $T_\text{QW} = 460\SI{}{\celsius}$ and $T_\text{GB} = 410\SI{}{\celsius}$ and (b) sample D, $T_\text{QW} = 410\SI{}{\celsius}$ and $T_\text{GB} = 370\SI{}{\celsius}$. Insets show $R_{xx}$ at a smaller range of $B$ to emphasize the onset of Shubnikov de Haas oscillations, denoted by an arrow.}
    \label{fig:fig2}
\end{figure}

In this work, we study the structural and transport properties of InAs near-surface quantum wells grown by molecular beam epitaxy (MBE). First, we present the evolution of the surface morphology of InAs without an In$_{0.81}$Al$_{0.19}$As capping layer for different graded buffer and quantum well growth temperatures, $T_\text{GB}$ and $T_\text{QW}$ respectively. We then discuss the use of an In$_{0.81}$Al$_{0.19}$As capping layer for enhanced electron mobility in InAs near-surface quantum wells. By studying the thickness dependence of the  In$_{0.81}$Al$_{0.19}$As cap on 2D electron density $n$ and mobility $\mu$, we characterize the In$_{0.81}$Al$_{0.19}$As surface properties and its effect on 2DEG transport. We find remarkable agreement between measured electron densities and those predicted by the Schrödinger and Poisson equations for a surface Fermi level pinning of $E_\text{pin} = \SI{40}{meV}$ for the In$_{0.81}$Ga$_{0.19}$As and $E_\text{pin} = \SI{309}{meV}$ for the In$_{0.81}$Al$_{0.19}$As.

A schematic of our heterostructure is shown in Fig.~\ref{fig:fig1}(a). Samples are grown on 2-inch epi-ready semi-insulating InP (100) wafers. All temperatures are measured by a thermocouple near the sample stage. The native oxide of the InP substrate is desorbed at a temperature of \SI{515}{\celsius} in an As atmosphere. Oxide removal is confirmed by a transition in the reflection high-energy electron diffraction pattern(s) from a (2$\times$4) to a c(4$\times$4) surface reconstruction. A \SI{50}{nm} In$_{0.53}$Ga$_{0.47}$As/In$_{0.52}$Al$_{0.48}$As superlattice of 10 periods is grown at a temperature of \SI{490}{\celsius}, followed by a \SI{100}{nm} In$_{0.52}$Al$_{0.48}$As layer. The \SI{800}{nm} In$_{x}$Al$_{1-x}$As step-graded buffer layer is grown with a digitally graded composition from $x=0.52$ to $x=0.81$ with \SI{50}{nm} steps of $\Delta x = 0.02$. The growth temperature of the graded buffer layer, $T_\text{GB}$, ranged from \SI{330}{\celsius} to \SI{410}{\celsius}. Two growth temperatures for the active region are used, $T_\text{QW} = \SI{410}{\celsius}$ and \SI{460}{\celsius}. A thin \SI{100}{nm} In$_{0.81}$Al$_{0.19}$As virtual substrate is utilized, followed by a $\delta$-doped Si layer with a density of 1.0$\times$10$^{12}$ cm$^{-2}$. We then grow a \SI{6}{nm} In$_{0.81}$Al$_{0.19}$As spacer layer, followed by a \SI{4}{nm} InAs channel confined by \SI{4}{nm} and  \SI{10}{nm} In$_{0.81}$Ga$_{0.19}$As bottom and top barriers. An In$_{0.81}$Al$_{0.19}$As capping layer of thickness $d$ is then grown in selected samples. The wafer is then cooled to $\SI{-15}{\celsius}$ for the deposition of \SI{10}{nm} of aluminum. 

In our study the graded buffer layer is grown colder than the oxide desorption temperature of InP and the growth temperature of the active region to promote the formation of misfit dislocations at layer step interfaces. We find that incremental changes to $T_\text{GB}$ can also substantially affect the formation and kinetics of these dislocations. Samples presented in this study are summarized in Table~\ref{table:table}. Samples A, B, and C are grown at $T_\text{QW} = \SI{460}{\celsius}$ with $T_\text{GB}$ = \SI{410}{\celsius}, \SI{400}{\celsius}, and \SI{390}{\celsius}, while Samples D, E, and F are grown at $T_\text{QW} = \SI{410}{\celsius}$ and $T_\text{GB}$ = \SI{370}{\celsius}, \SI{350}{\celsius}, and \SI{330}{\celsius}. In Fig.~\ref{fig:fig1}(b-g), we show $20\times\SI{20}{\micro m}$ atomic force microscopy (AFM) images of the aluminum surface of Samples A-F shown in alphabetical order. In all AFM images, one can see a two-dimensional crosshatched pattern, indicative of underlying misfit dislocations in the buffer layer. Relaxation along the $[110]$ and $[1\bar{1}0]$ directions is controlled by the nucleation and glide of $\alpha$ and $\beta$ dislocations respectively, and their different kinetics result in an anisotopic surface morphology~\cite{hudait_comparison_2004}. 

At higher temperatures (\SI{410}{\celsius}), the surface is prone to roughening in order to relax residual strain, promoting the Stranski–Krastanov growth mode. We find the root-mean-squared surface roughness, $Rq$, generally decreases with decreasing $T_\text{GB}$. Sample A has $Rq =\SI{5.85}{nm}$, dominated by large grain-like features. Sample B has a decreased surface roughness of $Rq =\SI{3.71}{nm}$ and is noticeably less grainy. In Sample C, we find a uniform, crosshatched morphology with occasional point defects, presumed to be due to Group V deficient growth. The roughness of this sample is $Rq =\SI{6.02}{nm}$ over the whole area and $Rq=\SI{1.46}{nm}$ in the area outlined by the dashed box which is near but excluding any point defects. The difference in $Rq$ in the two areas suggest that relaxation occurs in the areas surrounding defects.

Samples D, E and F are grown with lower $T_\text{GB} =$ 370, 350, and \SI{330}{\celsius} while also decreasing $T_\text{QW}$ from $460\SI{}{\celsius}$ to $410\SI{}{\celsius}$. We find $Rq$ are considerably lower than those of Samples A, B, and C, being $Rq =$ \SI{1.95}{nm}, \SI{1.70}{nm}, and \SI{1.88}{nm} respectively. The peak-to-peak heights are also considerably lower, improving from $\SI{33}{nm}$, $\SI{22}{nm}$ and $\SI{15}{nm}$ to around $\SI{10}{nm}$. Samples D, E, and F have a similar $Rq$ but their surface morphology differs qualitatively. We see the crosshatch pattern stemming from threading dislocations in the buffer layer becomes less noticeable between Samples D, E, and F, and in Sample F the crosshatch pattern is almost completely unnoticeable.



Since surface roughness creates electron scattering sites and directly affects electron transport \cite{Kaushini2018}, we conduct low-temperature (\SI{1.5}{\kelvin}) magnetotransport studies. We determine the 2D electron density $n$ and mobility $\mu$ of Samples A and D, representative of the two $T_\text{QW}$ regimes. Samples are measured in a van der Pauw configuration after the Al layers are etched using a wet chemical etchant (Transene Type D) in order to measure the transport properties of the 2DEG. Longitudinal and Hall resistances $R_{xx}$ and $R_{xy}$ as a function of out-of-plane magnetic field $B$ are shown in Fig.~\ref{fig:fig2}(a) and (b) for Samples A and D respectively. The two in-plane directions of the sample are $x$ and $y$ corresponding to the $[110]$ and $[1\bar{1}0]$ crystal directions. The growth temperatures $T_\text{QW}$ and $T_\text{GB}$ are annotated. One can see that Sample A has $\mu = \SI{1.49e4}{cm^2/Vs}$ while Sample D has $\mu = \SI{2.70e4}{cm^2/Vs}$. We observe well developed quantum Hall states with corresponding filling factors $\nu$ = 2, 4, and 6 for Sample A and $\nu$ = 3, 4, and 6 for Sample D. Samples A and D have similar 2D electron densities  $n=\SI{6.88e11}{cm^{-2}}$ and $n=\SI{7.81e11}{cm^{-2}}$ respectively. Shubnikov de Haas (SdH) oscillations are present in both samples. The insets of Fig.~\ref{fig:fig2}(a) and (b) show a zoomed in magnetic field range of the longitudinal resistance to focus on the onset of SdH oscillations, being 2.37 T for Sample A and 2.15 T for Sample D. The higher mobility of Sample D is associated with lower roughness and smoother surface morphology.

\begin{figure}
    \centering
    \includegraphics[width=0.45\textwidth]{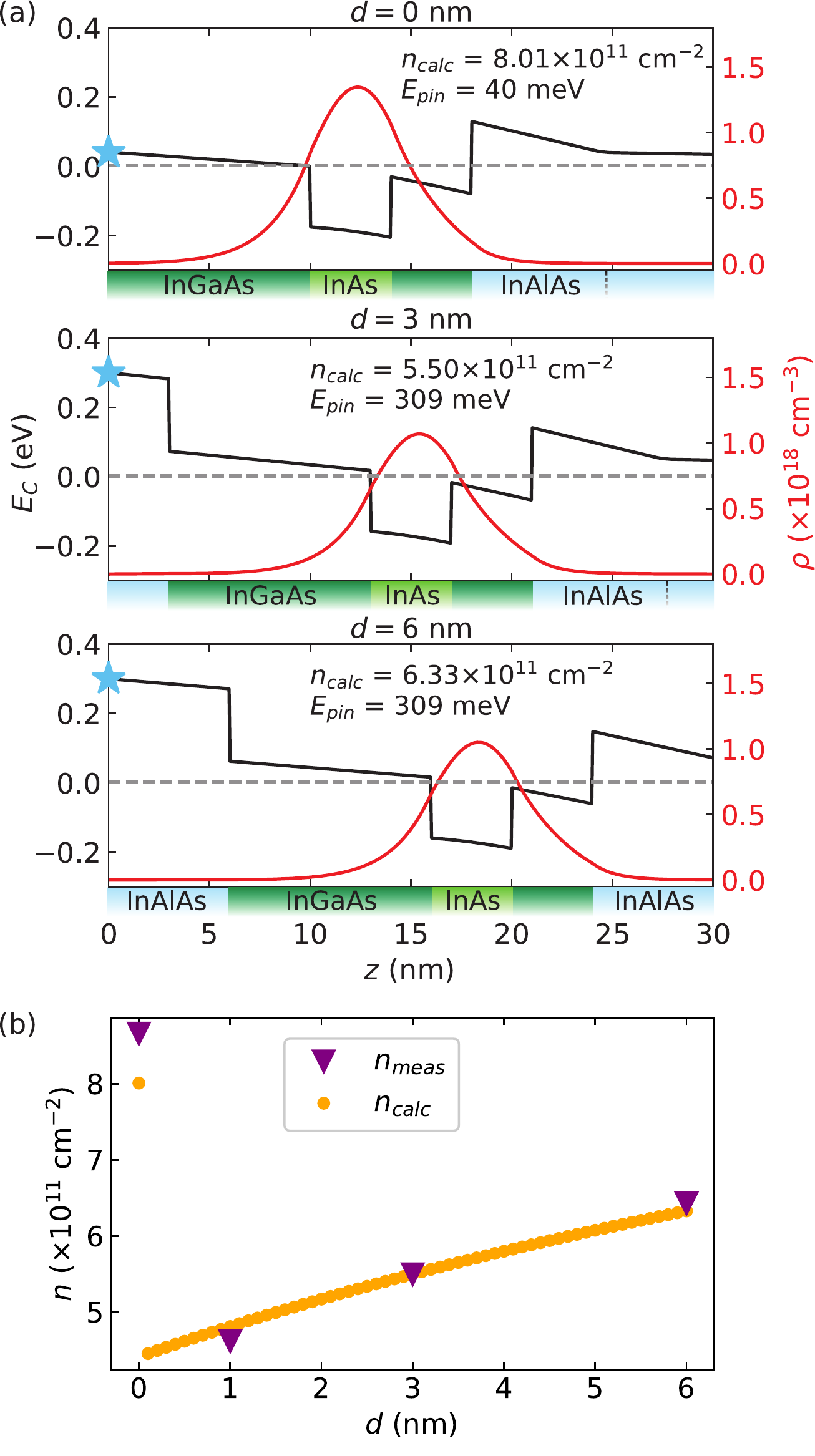}
    \caption{(a) Conduction band energy $E_\text{C}$ (black, left) and charge density $\rho$ (red, right) as a function of the distance from the surface $z$ for three different In$_{0.81}$Al$_{0.19}$As barrier thicknesses $d = $ \SI{0}{nm}, \SI{3}{nm}, and \SI{6}{nm}. Calculated density $n_\text{calc}$ for each case is shown. Calculations assume the conduction band energy at the surface is pinned to be an energy $E_\text{pin}$ (blue stars) above the Fermi level (gray dotted line). We use $E_\text{pin} =\SI{40}{meV}$ for the In$_{0.81}$Ga$_{0.19}$As top barrier and $E_\text{pin} = \SI{309}{meV}$ for the In$_{0.81}$Al$_{0.19}$As cappping layer. Layers of the underlying materials are noted on the $z$ axis, with labels for the given material corresponding to the color. (b) Measured electron densities (purple triangles) and calculated densities (yellow circles) as a function of the In$_{0.81}$Al$_{0.19}$As barrier thickness $d$ are shown.}
    \label{fig:fig3}
\end{figure}

It is evident that the transport properties of the near-surface InAs 2DEG are very sensitive to the top barrier material. InAs itself has a negative Fermi level pinning causing its conduction band to bend down at the surface, and electron mobilities are limited to below $\mu = \SI{10e4}{cm^2/Vs}$ \cite{tsui1970}. In order to alleviate the effects of surface scattering, one can utilize alloyed In$_x$Ga$_{1-x}$As top barriers to tune the surface Fermi level pinning \cite{Shabani2016}. However, an alternative barrier layer or additional capping layer such as In$_{0.81}$Al$_{0.19}$As to alter the surface conduction band edge, where a larger band offset and Schottky barrier may lead to better confinement of the 2DEG and in turn promote higher mobilities. This layer can not be too thick as the Schottky barrier quickly decouples the superconductor from the 2DEG within a few nanometers. 

We study a series of four samples with In$_{0.81}$Al$_{0.19}$As barriers of thicknesses $d=$ 0, 1, 3, and \SI{6}{nm} grown. These samples are grown with optimized growth temperatures as found in the previous section, being $T_\text{GB} = \SI{355}{\celsius}$ and $T_\text{QW} = \SI{410}{\celsius}$. We use the nextnano software to calculate the 2D electron densities $n_\text{calc}$, the 3D charge distribution $\rho$ and the conduction band energy $E_\text{C}$ by solving the Schrödinger and Poisson equations self-consistently~\cite{birner2007}. The band gaps for InAs, In$_{0.81}$Ga$_{0.19}$As, and In$_{0.81}$Al$_{0.19}$As used in the calculations are \SI{0.372}{eV}, \SI{0.520}{eV}, and \SI{0.880}{eV} respectively. The band offsets of InAs, In$_{0.81}$Ga$_{0.19}$As, and In$_{0.81}$Al$_{0.19}$As relative to InP are \SI{-0.670}{eV}, \SI{-0.550}{eV}, and \SI{-0.210}{eV} respectively. The depth $z$ is measured from the surface of the top layer  towards the substrate. The Fermi level lies at \SI{0}{eV}. Simulations assume a fully ionized sheet charge $n_\text{D} = \SI{1.0e12}{cm^{-2}}$ at $z=d+\SI{24}{nm}$.

We find that the surface conduction band energy is pinned to an energy $E_\text{pin}$ above the Fermi level being $E_\text{pin} = \SI{40}{\milli eV}$ for In$_{0.81}$Ga$_{0.19}$As and $E_\text{pin} = \SI{309}{\milli eV}$ for  In$_{0.81}$Al$_{0.19}$As. The resulting $E_\text{C}$ and $\rho$ for $d=\SI{0}{}$, $ \SI{3}{}$, and $\SI{6}{\nano m}$ are shown in Fig.~\ref{fig:fig3}(a) with layers of the heterostructure annotated below. The larger $E_\text{pin}$ of In$_{0.81}$Al$_{0.19}$As causes $E_\text{C}$ at $z = d + \SI{10}{nm}$ (the start of the InAs layer) to increase from $E_\text{C} = \SI{-172}{\milli eV}$ to $E_\text{C} = \SI{-148}{meV}$. This causes the density to decrease from $n_\text{calc} = \SI{8.01e11}{cm^{-2}}$ for $d=\SI{0}{nm}$ to $n_\text{calc} = \SI{5.50e11}{cm^{-2}}$ for  $d=\SI{3}{nm}$. The density then increases to $\SI{6.33e11}{cm^{-2}}$ for $d=\SI{3}{nm}$ due to $E_\text{C}$ at the InAs layer decreasing to \SI{-156}{meV}. Calculations of $n_\text{calc}$ for $d$ ranging from \SI{0.1}{nm} to \SI{6.0}{nm} in steps of \SI{0.1}{nm} are shown in Fig.~\ref{fig:fig3}(b). We also show measured densities $n_\text{meas}$ for samples with $d$ = 0, 1, 3 and \SI{6}{nm}. We find the calculations reasonably capture the trend in $n_\text{meas}$. At $d =\SI{1}{nm}$, the density decreases from $n_\text{meas} = 8.66 \times 10^{11}$ cm$^{-2}$ to $n_\text{meas} = 4.62 \times 10^{11}$ cm$^{-2}$, due to the abrupt change in $E_\text{pin}$ from the In$_{0.81}$Ga$_{0.19}$As surface to the In$_{0.81}$Al$_{0.19}$As surface. The density then increases as $d$ increases, with $n_\text{meas} = 6.43 \times 10^{11}$ cm$^{-2}$ for $d=\SI{6}{nm}$.

\begin{figure}
    \centering
    \includegraphics[width=0.5\textwidth]{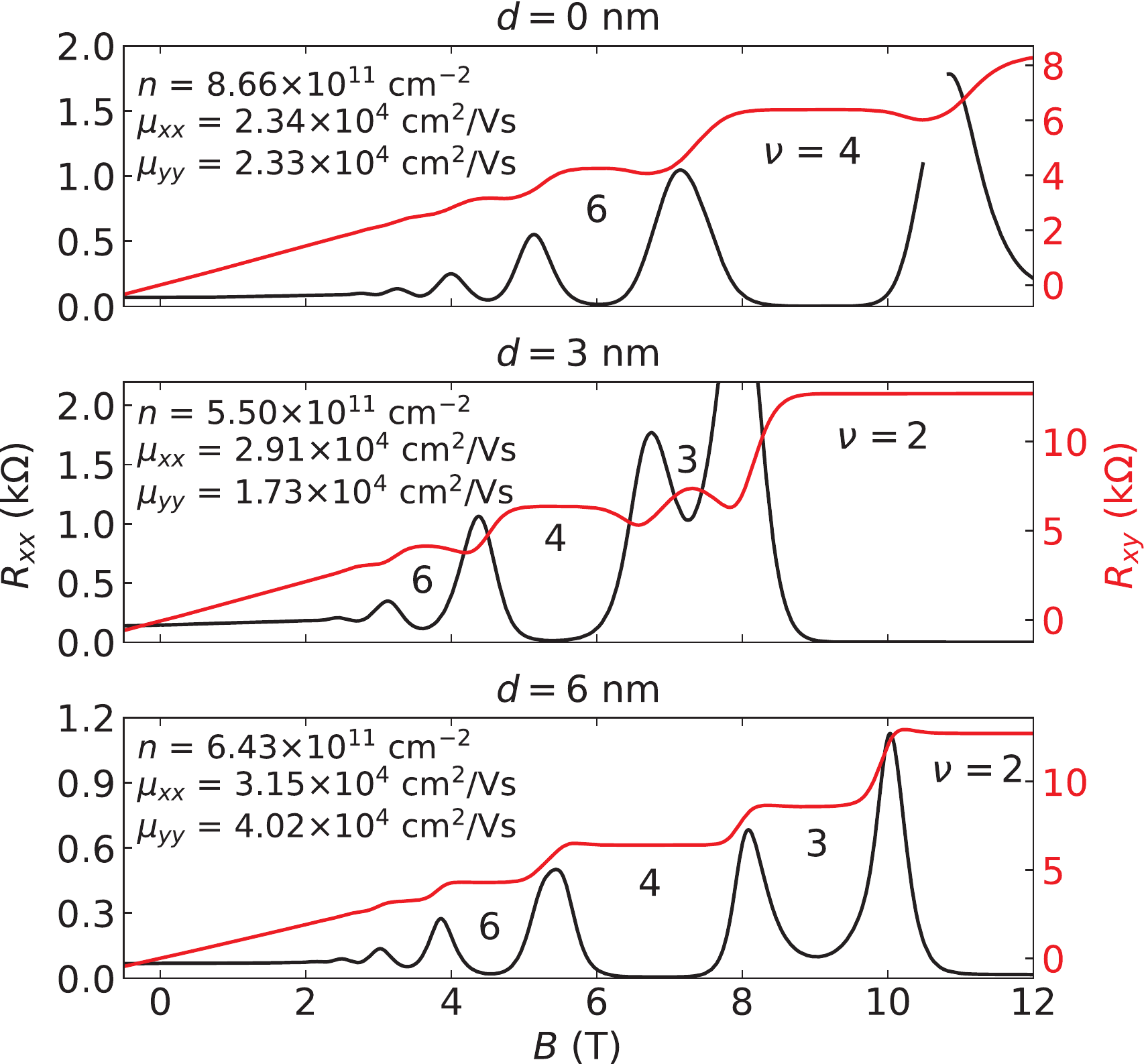}
    \caption{a) Longitudinal resistance $R_{xx}$ (black) and transverse resistance $R_{xy}$ (red) as a function of magnetic field $B$ for three different samples, with $d = $ \SI{0}{nm}, \SI{3}{nm}, and \SI{6}{nm}. The measured 2D electron density $n$ and mobilities $\mu_{xx}$ and $\mu_{yy}$ are shown, as well as the positions of different filling factors $\nu$.}
    \label{fig:fig4}
\end{figure}

Fig.~\ref{fig:fig4} shows magnetotransport data for three samples with $d$ = 0, 3, and \SI{6}{nm}. At the introduction of a \SI{3}{nm} In$_{0.81}$Al$_{0.19}$As cap $\mu$ initially decreases. At a thickness of \SI{6}{nm}, the electron mobility $\mu=\SI{3.59e4}{cm^2/Vs}$ is improved over $\mu$ measured for In$_{0.81}$Ga$_{0.19}$As capped samples in this report. In additon the Zeeman split filling factor $\nu$ = 3 is emerging. We note that the anisotropy of $\mu$ is due to the asymmetry of dislocations along the $x$ and y directions ($[110]$ and $[1\bar{1}0]$).

We estimate the surface Fermi level pinning of In$_{0.81}$Al$_{0.19}$As by fitting $n_\text{meas}$ to $n_\text{calc}$ as a function of $d$. We measure the Fermi level pinning position of In$_{0.81}$Al$_{0.19}$As to be \SI{309}{\milli eV} at a composition of $x=0.81$. This is consistent with expected values for the Schottky barrier of In$_{0.81}$Al$_{0.19}$As with metals using standard extrapolation methods \cite{adachi2009_schottky, chyi1994, gueissaz1992, sadwick1991}. 


In summary, we have detailed improvements of surface morphology and electron mobility in the growth of InAs near-surface quantum wells. Our work shows the importance of combined structural and transport optimization and the role of surface Fermi level pinning. Using an In$_{0.81}$Al$_{0.19}$As capping layer, one can screen the 2DEG from surface scattering sites, enhancing electron mobilities as well as the presence of odd integer quantum Hall states. We find that Schrödinger-Poisson calculations of the electron density and conduction band energy accurately describe the trend in measured 2D electron densities for samples with 1~nm and above In$_{0.81}$Al$_{0.19}$As capping layer thicknesses. Using measured densities, we estimate the surface Fermi level pinning of In$_{0.81}$Al$_{0.19}$As to be \SI{309}{\milli eV} . 


\section{Acknowledgements}

The authors acknowledge support by DARPA TEE award no. DP18AP900007US, US Army Research Office agreement W911NF1810067. W. F. S. acknowledges funding from the NDSEG Fellowship. We are grateful to Ido Levy and Patrick J. Strohbeen for useful discussions.

\bibliography{References_Shabani_Growth}
\pagebreak

\end{document}